\pdfoutput=1
\documentclass[%
twocolumn,
showpacs,preprintnumbers,
 amsmath,amssymb,
 aps,
 superscriptaddress,
pra,
floatfix,
]{revtex4-1}

\usepackage{graphicx}
\usepackage{dcolumn}
\usepackage{bm}
\usepackage{hyperref}
\usepackage[utf8]{inputenc}
\usepackage{amsmath}
\usepackage[amssymb]{SIunits}
\usepackage{amsfonts}
\usepackage{amssymb}
\usepackage{subfigure}
\usepackage{mathtools}
\usepackage{dsfont}
\usepackage{color}
\usepackage{isotope}
\usepackage{calc}
\usepackage{braket}

\begin{document}


\title{Strong magnetic coupling of an inhomogeneous NV ensemble to a cavity}

\author{K. Sandner}
\affiliation{Institute for Theoretical Physics, Universität Innsbruck, Technikerstrasse 25, 6020 Innsbruck, Austria}
\author{H. Ritsch}
\affiliation{Institute for Theoretical Physics, Universität Innsbruck, Technikerstrasse 25, 6020 Innsbruck, Austria}

\author{R. Amsüss}
\author{Ch. Koller}
\author{T. Nöbauer}
\author{S. Putz}
\author{J. Schmiedmayer}
\author{J. Majer}
\affiliation{Vienna Center for Quantum
Science and Technology,
Atominstitut, TU-Wien,
1020 Vienna, Austria}

\date{\today}

\begin{abstract}
We study experimentally and theoretically a dense ensemble of negatively charged
nitrogen-vacancy centers in diamond coupled to a high $Q$ superconducting coplanar
waveguide cavity mode at low temperature. The nitrogen-vacancy centers are modeled as
effective spin one defects with inhomogeneous frequency distribution. For a large enough
ensemble the effective magnetic coupling of the collective spin dominates the mode losses
and inhomogeneous broadening of the ensemble and the system exhibits well resolved normal
mode splitting in probe transmission spectra. We use several theoretical approaches to
model the probe spectra and the number and frequency distribution of the spins. This
analysis reveals an only slowly temperature dependent q-Gaussian energy distribution of
the defects with a yet unexplained decrease of effectively coupled spins at very low
temperatures below $\unit{100}{\milli\kelvin}$. Based on the system parameters we predict
the possibility to implement an extremely stable maser by adding an external pump to the
system.        


\end{abstract}

\maketitle

\section{Introduction}
Systems of spin ensembles coupled to a cavity mode are considered a promising physical
realization for processing and storage of quantum information~\citep{phillips2001storage},
ultrasensitive high resolution magnetometers~\cite{taylor2008high} or localized field probes. Collective
magnetic coupling of a large ensemble to the field mode allows to reach the strong
coupling regime, even if a single particle is hardly coupled. Implementations based on
superconducting coplanar waveguide (CPW) resonators provide an interface of the spin ensemble
with processing units for superconducting qubits. Here the spins can serve as a quantum
memory or as a bridge to optical readout and communication~\citep{verdu2009strong}. 

Different types of ensembles were proposed for this setup, ranging from clouds of
ultracold atoms~\citep{verdu2009strong} over polar molecules~\citep{rabl2006hybrid} to
solid state systems like rare-earth spin ensembles~\citep{bushev2011ultralow} or color centers
in diamond~\citep{kubo2010strong,amsuss2011cavity}. Here we focus on the negatively
charged nitrogen-vacancy defects in diamond (NV). Those are naturally present in diamond
but can also be readily engineered with very high densities, still maintaining long
lifetimes and slow dephasing in particular at low temperatures of $T<\unit{1}{\kelvin}$. In the optical
domain, they are extremely stable and very well studied since many years~\citep{jelezko2006single}. 

The magnetic properties of the relevant defect
states can be conveniently modeled by effective independent spin 1 particles, where the
effective local interaction of the electrons within the defect shifts the $(m_{S}=\pm 1)$
states with respect to the $m_{S}=0$ state~\cite{childress2006coherent}. On the one hand, this coupling
provides the desired energy gap in the $\unit{3}{\giga\hertz}$ regime, but on the other hand,
as a consequence of local variations of the crystal field, this shift exhibits a
frequency distribution leading to an inhomogeneous broadening of the ensemble. The
inhomogeneities are thought to be predominantly caused by crystal strain and excess
nitrogen, which is not paired with a neighboring vacancy~\citep{acosta2009diamonds}. 

While for a perfectly monochromatic ensemble of $N$ particles in a cavity, strong coupling
simply requires an effective coupling $g_{\text{single}}\sqrt{N}$ larger than the cavity
and spin decay rates, not only the width, but also the details of the inhomogeneous
distribution are known to strongly influence the dynamic properties of the real world
system~\citep{houdre1996vacuum,diniz2011strongly,kurucz2011spectroscopic}. In particular a
Gaussian or a Lorentzian distribution of equal half width, lead to different widths and
magnitudes of the vacuum Rabi splitting. Only above a critical coupling strength the
rephasing via common coupling to the cavity mode will prevent dephasing of the collective
excitation and lead to a well resolved vacuum Rabi splitting.
 
In our theoretical studies of this system we use different approximation levels to analyze the central physical
effects present in an inhomogeneously broadened system, as they are observed in the
measurements. While many qualitative features can be readily understood from a simple,
coupled damped oscillator model with an effective linewidth, a detailed understanding of the observed frequency
shifts and coupling strengths in the experiments requires more sophisticated modeling of
the energy distributions and dephasing mechanisms. In particular the observed temperature
dependence relies on a finite temperature master equation treatment of a collective spin
with proper dephasing terms. This is compared to the experimental values of up to ($N\approx 10^{12}$)
particles as presented already in~\citep{amsuss2011cavity}.
  
The paper is organized as follows: The general properties of the system are introduced in
Sec.~\ref{sec:general}.  A first approximative treatment using coupled harmonic
oscillators is shown in Sec.~\ref{subsec:HO}. In Sec.~\ref{subsec:gamma_p} we incorporate
the inhomogeneous frequency distribution of the NVs via an extra decay of the polarization
of the ensemble. In this context we also analyze the effects of thermal excitations in the
spin ensemble and in the cavity. Finally, in Sec.~\ref{subsec:resolvent} we interpret
our measurements using the resolvent formalism in order to extract the exact form of the
inhomogeneity~\citep{kurucz2011spectroscopic}. The prospect of implementing a narrow
bandwidth transmission line micro-maser using an inhomogeneously broadened ensemble is
discussed in Sec.~\ref{sec:maser}.

\section{General system properties and experimental implementation}\label{sec:general}

The ground state of the NV center is a spin triplet $(S=1)$, where the $m_{S}=\pm 1$ states are split
from the $m_{S}=0$ state by about $\unit{2.88}{\giga\hertz}$ at zero magnetic
field~\citep{acosta2009diamonds}. In addition the
degeneracy between the $m_{S}=\pm 1$ states is lifted due to the broken $C_{3v}$ symmetry. Applying a homogeneous 
magnetic field enables Zeeman tuning of the $m_{S}=\pm 1$ states, so that the $m_{S}=0\rightarrow \pm1$ 
transitions can be tuned selectively into resonance with the CPW cavity. 

Note that here the Zeeman tuning is also varying with the NV centers orientation relative to the applied field direction.
In NV center ensembles all the four symmetry allowed orientations of the NV main axis are found and in general
each orientation will enclose a different angle with the magnetic field and will be shifted by a different amount.

However, if the magnetic field direction is oriented within the (001) plane of the diamond, always two of four orientations
exhibit the same angle with the field and at special angles all four are tuned by the same amount. 

In general we therefore have to distinguish between the $m_{S}^{\text{I,II}}=\pm 1$ states of
subensemble I and II. The Hamiltonian for one NV subensemble can be written as
\begin{align}
	\operatorname{H_{\text{NV}}}=\tilde{g} \mu_{B} \mathbf{B} \cdot \mathbf{S} + D S_z^2 + E (S_x^2-S_y^2),
	\label{eq:HNV}
\end{align}
where the first term describes the Zeeman effect with $\tilde{g}=2$ for an NV center and $\mu_B$
being the Bohr magneton. The second term denotes the zero-field splitting with
$D=\unit{2.88}{\giga\hertz}$ and typical strain-induced $E$ parameters of several MHz. The presence of
non-zero E parameters in large NV ensembles causes a mixing of the pure Zeeman states for
low magnetic field values which are denoted as $\ket{0,\pm}$. As the magnetic field
amplitude is increased, the Eigenstates are again well approximated by pure Zeeman states.

The transition frequencies $\omega_{\pm}^{\text{I,II}}$ between the eigenstates
$\ket{0}_{\text{I,II}}$ and $\ket{\pm}_{\text{I,II}}$ are depicted in
Fig.~\ref{fig:HOmodel}(b).

\subsection{Theoretical model}\label{subsec:theory}

Assuming that only the transition $\ket{0}_{\text{I}}\rightarrow \ket{-}_{\text{I}}$ with
frequency $\omega_{-}^{\text{I}}$ is in or close to resonance with the cavity mode, we
can reduce the description of a single  NV to a 2-level system. The effects caused by the presence of the 
far detuned $ \ket{+}_{\text{I}}$ state can be included in a constant effective frequency shift as addressed in Sec.~\ref{subsec:HO}.  
We thus approximate the composed system of cavity and ensemble with the Tavis-Cummings Hamiltonian
\begin{align}
	\operatorname{H}_{\text{TC}}=\omega_{c}a_{c}^{\dagger}a_{c}+\frac{1}{2}\sum^{N}_{j}\omega_{j}\sigma_{j}^{z}+\sum^{N}_{j}\left(g_{j}\sigma_{j}^{+}a_{c}+\text{H.c.}\right)\
	,
	\label{eq:Hamiltonian_no_HP}
\end{align}
where $\hbar=1$. 
The first two terms describe the unperturbed energies of the cavity, with frequency
$\omega_{c}$ and creation operator $a_{c}^{\dagger}$, and of the $N$ ensemble spins using
the usual Pauli spin operators $\left(\left[ \sigma_{i}^{+},\sigma_{j}^{-}
\right]=\sigma_{i}^{z}\delta_{ij}\right)$. Each spin can have a different frequency
$\omega_{j}$ which is statistically spread around the center frequency
$\omega_{-}^{\text{I}}$. The third term describes the coupling to the cavity with
individual strength $g_{j}$. Assuming that the ensemble spins are confined to a volume
small compared to the wavelength, the ensemble will interact collectively with the mode.
In the case of few excitations, the ensemble behaves like a harmonic oscillator which
couples to the mode with the collective coupling strength $\Omega$. The collective
coupling strength is given by $\Omega=\sqrt{\sum_{j}^{N}\left| g_{j}\right| ^{2}}$ which
for identical $g_{j}=g_{\text{single}}=g$ gives $\Omega=g\sqrt{N}$. To include the probe field of the cavity
we add another term $\operatorname{H}_{p}=\mathrm{i} \left(\eta a_{c}^{\dagger
}\mathrm{e}^{-\mathrm{i}\omega_{p} t}-\eta^{*}a_{c}\mathrm{e}^{\mathrm{i}\omega_{p} t}
\right)$ to Eq.~\ref{eq:Hamiltonian_no_HP}. 

Effects of the coupling to a finite temperature bath are
analyzed in Sec.\ref{subsec:gamma_p}, where we study the master equation of the
ensemble-cavity system.

For an ensemble with identical frequencies and coupling strengths we find that we can
reach the regime of coherent oscillations between the cavity and the ensemble spins if the
collective coupling $\Omega$ dominates the linewidth of the cavity $\kappa$ and the decay
rate of the single spin $\gamma_{\text{hom}}$. This is commonly known as strong coupling
regime. In case of an ensemble with inhomogeneous frequency distribution it is not
immediately clear under which conditions we can observe the avoided crossing. Here we
address the influence of the width and form of the inhomogeneous frequency distribution
on the avoided crossing.

\subsection{Experimental setup and parameters}\label{subsec:experiment}

The experimental set-up has been described already in detail in \cite{amsuss2011cavity} and is
explained here only in brevity in the following. The heart of the experiment is a $\lambda /2$
superconducting coplanar waveguide (CPW) resonator with a center frequency of
$\unit{2.7}{\giga\hertz}$ that is cooled down to $\unit{20}{\milli\kelvin}$ in a dilution
refrigerator. In order to couple NV defect centers to the microwave field in the cavity, a (001)-cut single crystal diamond is placed in
the middle of the resonator, where the oscillating magnetic field exhibits an antinode. A
meander geometry of the resonator ensures that a large fraction of the magnetic mode volume is covered by
spins. The high-pressure high-temperature diamond chosen in this setup contains an NV concentration of about $6$~ppm,
which corresponds to an average separation of about $\unit{10}{\nano\meter}$.

A 2-axis Helmholtz coil configuration creates a homogeneous static magnetic field oriented with an
arbitrary direction within the (001) plane of the diamond, which is at the same time
parallel to the resonator chip surface. Since perpendicular magnetic field components with
respect to the resonator surface would shift the resonance frequency of the cavity, great
care was taken during the alignment of the set-up. With the diamond on top of the chip
the cavity quality factor is $Q=3200$.

In a typical experiment, we first set the magnetic field direction and amplitude and then
measure the microwave transmission through the cavity with a vector network analyzer.

\section{Cavity transmission spectra for inhomogeneous ensembles}\label{sec:approaches}

As generic experiment to test and characterize the system properties we analyze the
weak field probe transmission spectrum through the resonator in the weak field limit,
where the number of excitations is negligible compared to the ensemble size. Hence, at
least at low temperatures, we can largely ignore saturation effects and apply 
various simplified theoretical descriptions to extract the central system parameters. 
In fact as we deal with more than $N>10^{11}$ spins in a transition frequency range of
about $\unit{10^{7}}{\hertz}$,
we have several thousand spins per Hz frequency range and up to a million spins within the homogeneous width 
of at least several kHz. Hence theoretical modelling as a collection of effective oscillators should provide an
excellent model basis. This has to be taken with care at higher temperatures $k_{B}T \approx \hbar \omega_c$,
when we have a significant fraction of the particles excited. 
 
\subsection{Coupled collective oscillator approximation}\label{subsec:HO}

A photon that enters the cavity will be absorbed into a symmetric excitation of the ensemble
spins with a weight given by the individual coupling strength. Subsequently the broad
distribution of the spin frequencies induces a relative dephasing of this excitation so
that the backcoupling to the cavity mode is suppressed. As spontaneous decay
($T_{1}\approx \unit{44}{\second}$) is
negligibly slow~\citep{amsuss2011cavity}, the decay rate of the polarization of the ensemble is just proportional
to this dephasing and thus the inhomogeneous width of the spins.  

As the number of photons entering the ensemble is very small compared to $N$ in a first
model, we simply approximate the ensemble as a harmonic oscillator with frequency
$\omega_{a1}=\omega_{-}^{\text{I}}$ and an effective large width $\gamma$ that mimics the
inhomogeneity. The ensemble oscillator is coupled to the cavity with frequency
$\omega_{c}$ and decay rate~$\kappa$. 

This very simplified model already allows us to study the coupling dynamics of the
collective energy levels of the broadened ensemble and the cavity mode, and in particular
the avoided crossing of the energy levels, when the relative energies are varied.

To keep the model as simple as possible but still grasping the essential properties, the
presence of the other offresonant spin ensembles at $\omega_{+}^{\text{I}}$,
$\omega_{-}^{\text{II}}$ and $\omega_{+}^{\text{II}}$ is modelled by additional
oscillators with frequency $\omega_{aj}$, ($j=2,3,4$) and equal decay rate $\gamma$.
 
For a small probe field injected into the cavity with the frequency $\omega_{p}$ the
corresponding equations in a frame rotating with $\omega_{p}$ then read:
\begin{align}
	\frac{\mathrm{d}}{\mathrm{d}t}\left< a_{c} \right>&=-\left(\kappa +\mathrm{i}
	\Delta_{c}\right)\left< a_{c} \right>
	-\mathrm{i}g\sum_{j=1}^{4}N_{j}\left< \sigma_{j}^{-}
	\right>+\eta
	\label{eq:2HO1}\\
	\frac{\mathrm{d}}{\mathrm{d}t}\left< \sigma_{j}^{-} \right>&=-\left(
	\frac{\gamma}{2}+\mathrm{i}\Delta_{aj} \right)\left< \sigma_{j}^{-} \right>-\mathrm{i}g\left< a_{c} \right>
	\label{eq:2HO2}\ ,
\end{align}
with probe amplitude $\eta$, $\Delta_{c}=\omega_{c}-\omega_{p}$ and
$\Delta_{aj}=\omega_{aj}-\omega_{p}$. From Eqs.~\ref{eq:2HO1} and \ref{eq:2HO2} we can calculate
the steady state field in the cavity $\left< a_{c} \right>_{\text{st}}$, which can be
written as
\begin{align}
	\left< a_{c} \right>_{\text{st}}=\frac{\eta}{\kappa+\Gamma_{a}+\mathrm{i}\left(
	\Delta_{c}-U_{a}
	\right)+\frac{g^{2}N_{1}}{\gamma/2+\mathrm{i}\Delta_{a1}}}\ ,
	\label{eq:a_st}
\end{align}
where
\begin{align}
	\Gamma_{a}=\sum_{j=2}^{4}\frac{g^{2}N_{j} \gamma/2}{\left( \gamma/2
	\right)^{2}+\Delta_{aj}^{2}}\ \ \text{and}\ \
	U_{a}=\sum_{j=2}^{4}\frac{g^{2}N_{j} \Delta_{aj}}{\left( \gamma/2
	\right)^{2}+\Delta_{aj}^{2}}\ .
	\label{eq:Gamma_and_U}
\end{align}
In the absence of the offresonant levels ($N_{2}=N_{3}=N_{4}=0$) we recover the situation,
where for $\omega_{a1}=\omega_{c}$ and $g\sqrt{N_{1}} > (\gamma/2-\kappa)/2$ we find two
normal modes split by $2 \sqrt{g^{2}N_{1}-(\gamma/2-\kappa)^{2}/4}$. The offresonant
transitions make the situation more complex. As it can be seen in Eq.~\ref{eq:a_st}, they
induce a shift $U_{a}$ of the cavity frequency and increase the decay rate of the
cavity by $\Gamma_{a}$. Both, shift and additional decay rate, depend on the probe
frequency $\omega_{p}$. For the parameters in our measurement we find that $\Gamma_{a}$ is
negligible compared to $\kappa$. We further approximate $U_{a}$ by setting
$\omega_{p}=\omega_{c}$ (as the scan range of $\omega_{p}$ around $\omega_{c}$ is small compared
to $\omega_{aj}$ ($j=2,3,4$)). We obtain the model of two coupled
oscillators, where one of them has been shifted by the offresonant transitions. As a
function of $\omega_{p}$ and $\omega_{a1}$ (tuned by the magnetic field), $\left| \left< a_{c}
\right>_{\text{st}}\right|^{2} $ shows an avoided crossing. Although the
offresonant transitions are shifted by the magnetic field as well, the effect in
$U_{a}$ is negligibly small, so that we assume that $U_{a}$ is constant.

\begin{figure*}[t]
	\centering
	\includegraphics[width=\textwidth]{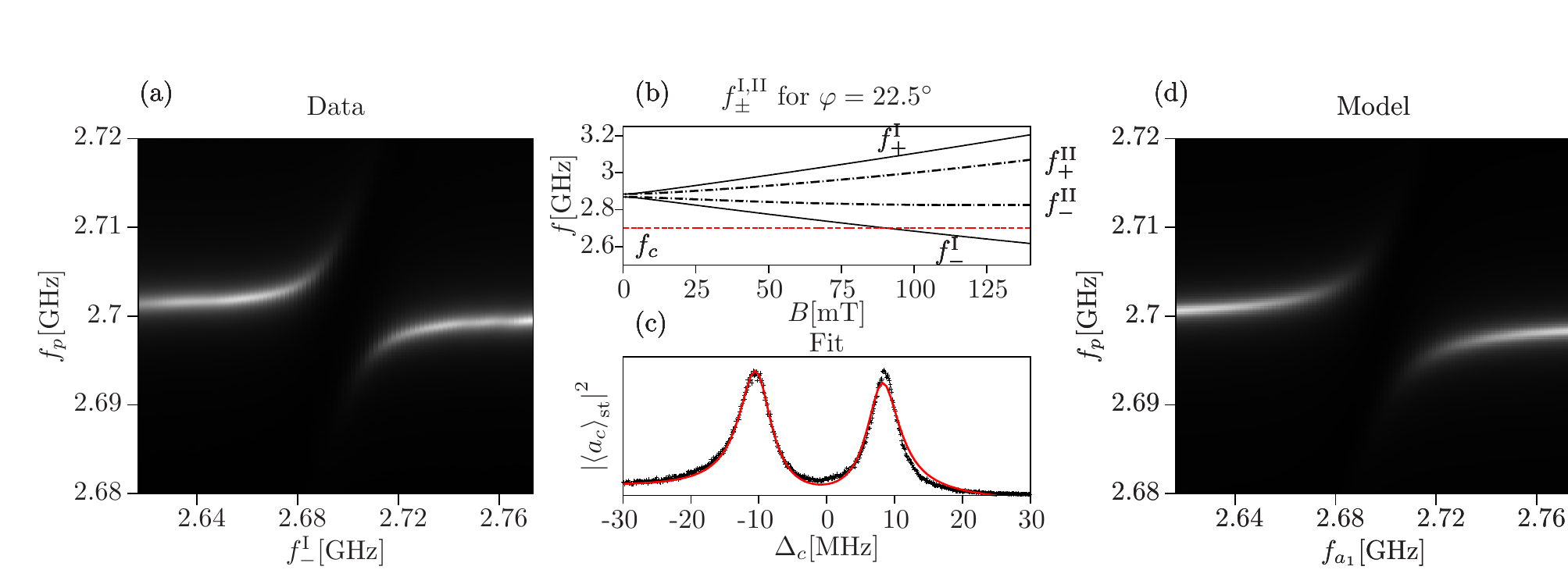}
	\caption{(a) (Color online) Measured transmission as a function of
	$f_{-}^{\text{I}}=\omega_{-}^{\text{I}}/(2\pi)$ tuned by the magnetic field and pump
	frequency $f_{p}=\omega_{p}/(2\pi)$. (b) Transition frequencies
	$f_{\pm}^{\text{I,II}}=\omega_{\pm}^{\text{I,II}}/(2\pi)$ as a function of the magnetic
	field for a field angle of $\varphi=22.5^{\degree}$. The frequency of the cavity
	$f_{c}=\omega_{c}/\left( 2\pi \right)$ is denoted by the (red) dashed
	line. (c) Fit of
	$\left| \left< a_{c} \right>_{\text{st}}\right|^{2}$ to the transmitted signal close to
	the resonance. We obtain the decay rate
	$\gamma=\unit{10.92}{\mega\hertz}$.
	(d) $\left| \left< a_{c} \right>_{\text{st}}\right|^{2}$ as a function of $f_{a_{1}}$ and
	$f_{p}$ incorporating the parameters obtained from the fit.}
	\label{fig:HOmodel}
\end{figure*}

We show the measured signal at the avoided crossing in Fig.~\ref{fig:HOmodel}(a). By
fitting $\left| \left< a_{c} \right>_{\text{st}}\right|^{2} $ to the normal mode splitting
we can deduce the inhomogeneous width $\gamma$ and the effective coupling $g\sqrt{N}$. The
$Q$ factor of our cavity corresponds to $\kappa/(2\pi)=\unit{0.4}{\mega\hertz}$, then
$2\kappa$ is the full width at half maximum (FWHM) of the cavity resonance. An exemplary fit
result is shown in Fig.~\ref{fig:HOmodel}(c). We have included a linear term in the fit to
account for a background in the data.  In Fig.~\ref{fig:HOmodel}(d) we plot $\left| \left<
a_{c} \right>_{\text{st}}\right|^{2}$ using the obtained parameters.

\begin{figure}[t]
	\centering
	\includegraphics[width=0.5\textwidth]{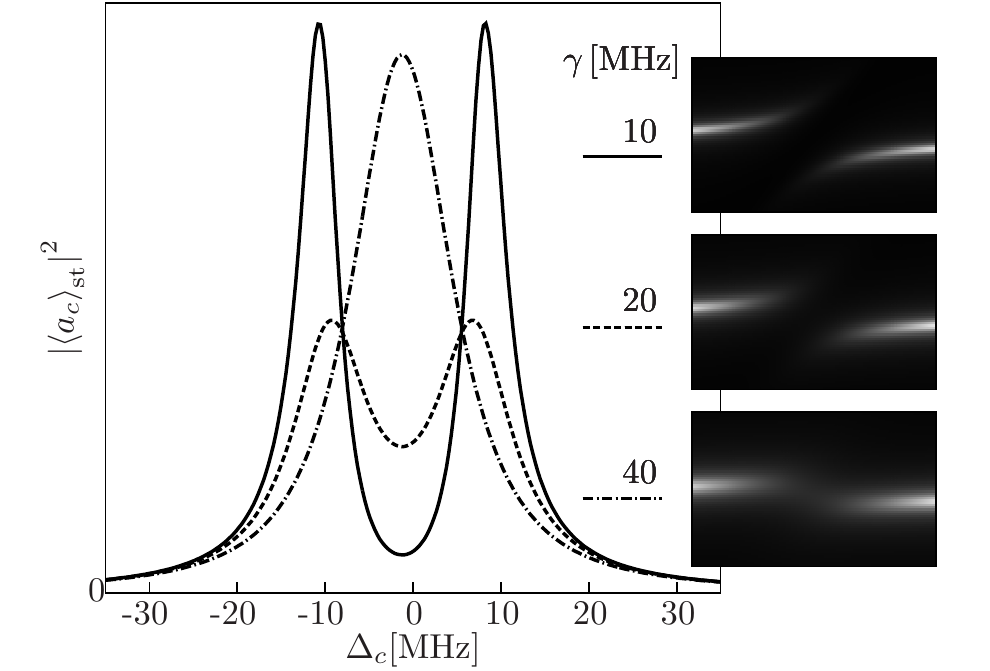}
	\caption{Simulation of the normal mode splitting at resonance
	($\omega_{c}-U_{a}=\omega_{a1}$) for different values of $\gamma$. Finally for large
	values of $\gamma$ the doublet
	is not resolved any more and the avoided crossing (sketched on the right) becomes
	blurred.}
	\label{fig:HOmodel_gamma}
\end{figure}

From the fit we deduce a decay rate $\gamma=\unit{10.92}{\mega\hertz}$ of the ensemble
oscillator and an effective coupling of $g\sqrt{N}=\unit{9.51}{\mega\hertz}$.
To demonstrate the effect of the ensemble oscillator width $\gamma$
on the normal mode splitting, we plot $\left| \left< a_{c} \right>_{\text{st}}\right|^{2}$
for $\omega_{c}-U_{a}=\omega_{a1}$ and different values of $\gamma$ in
Fig.~\ref{fig:HOmodel_gamma}. 

Treating the ensemble as a broad harmonic oscillator with decay rate $\gamma$
corresponds to the assumption that the frequency distribution of the real ensemble is a
Lorentzian distribution, given that all spins couple with equal
strength~\cite{kurucz2011spectroscopic}.

\subsection{Polarization decay and collective coupling at finite temperature}\label{subsec:gamma_p}

Despite cooling to very low temperatures is possible in the experiments, it is still important and instructive 
to study the role of thermal excitations in the system. In contrast to previous models based on virtually
zero $T$ atomic ensembles, the NV centers are in thermal contact with the chip at small
but finite $T$. 
We will now investigate how sensitive the system reacts on thermal fluctuations. 

At this point we include any shifts caused by the off-resonant ensembles in an effective
detuning and concentrate on the collective coupling between the cavity at $\omega_{c}$
and the near resonant ensemble centered around $\omega_{-}^{\text{I}}$. For simplicity we
assume equal coupling $g$ for all spins to get
\begin{align}
	\operatorname{\tilde{H}}_{\text{TC}}=\omega_{c}a_{c}^{\dagger}a_{c}+\frac{1}{2}\omega_{-}^{\text{I}}\sum^{N}_{j}\sigma_{j}^{z}+g\sum^{N}_{j}\left(\sigma_{j}^{+}a_{c}+\text{H.c.}\right)\
	.
	\label{eq:Hamiltonian_TC_gamma_p}
\end{align}
The effects of unequal coupling of the spins is addressed in~\cite{braun2011superradiance}.

To include thermal excitations of the mode and the ensemble we have to add standard Liouvillian
terms to the dynamics and study the corresponding master equation of the reduced cavity
ensemble system~\cite{carmichael1993open}, where the inhomogeneous width of the ensemble
is still simply approximated by an effective dephasing term $\gamma_p$ for the
polarization. In this model decay $\gamma_{\text{hom}}$ and dephasing $\gamma_p$ are
described by separate quantities, which already should improve the model. The master
equation reads
\begin{align}
	\frac{\mathrm{d}}{\mathrm{d}t}\rho=\frac{1}{\mathrm{i}}\left[
	\operatorname{\tilde{H}}_{\text{TC}}+\operatorname{H}_{p},\rho
	\right]+\mathcal{L}\left[ \rho \right]\ ,
	\label{eq:Master}
\end{align}
where
\begin{align}
	&\mathcal{L}\left[ \rho \right]=\kappa\left( \bar{n}\left(T, \omega_{c} \right)+1 \right)\left(2a_{c}\rho
	a_{c}^{\dagger }- a_{c}^{\dagger
	}a_{c}\rho-\rho a_{c}^{\dagger } a_{c}  \right)\notag\\
	&+\kappa\bar{n}\left(T, \omega_{c} \right)\left( 2a_{c}^{\dagger }\rho a_{c} -a_{c} a_{c}^{\dagger}\rho-\rho
	a_{c} a_{c}^{\dagger }
        \right)\notag\\         
				&+\frac{\gamma_{\text{hom}}}{2}\left( \bar{n}\left( T,\omega_{-}^{\text{I}} \right)+1
				\right)\sum_{j=1}^{N}\left(2\sigma_j^{-}\rho
				\sigma_j^{+}-\sigma_j^{+}\sigma_j^{-}\rho -\rho
        \sigma_j^{+}\sigma_j^{-} \right)\notag\\
				&+\frac{\gamma_{\text{hom}}}{2}\bar{n}\left( T,\omega_{-}^{\text{I}} \right)\sum_{j=1}^{N}\left(2\sigma_j^{+}\rho
				\sigma_j^{-}-\sigma_j^{-}\sigma_j^{+}\rho -\rho
        \sigma_j^{-}\sigma_j^{+} \right)\notag\\
				&+\frac{\gamma_{p}}{2}\sum_{j=1}^{N}\left( \sigma_{j}^{z}\rho
				\sigma_{j}^{z}-\rho \right)\ .
     \label{eq:Liou}
\end{align}
The first two lines of Eq.~\ref{eq:Liou} describe the coupling of the cavity to the
bath, while the next two lines include the coupling of the ensemble to the bath. The
number of thermal excitations at temperature $T$ and frequency $\omega$ is denoted by
$\bar{n}\left( T,\omega \right)$. The term
in the last line introduces nonradiative dephasing at a rate $\gamma_{p}$ of the spins and
thereby models the inhomogeneity.

Based on the master equation we can derive a hierarchic set of equations for various
system expectation values starting with
\begin{align}
	\frac{\mathrm{d}}{\mathrm{d}t}\left< a_{c}
	\right>&=\operatorname{Tr}\{a_{c}\frac{\mathrm{d}}{\mathrm{d}t}\rho\}\notag\\
	&=-\left( \kappa+\mathrm{i}\Delta_{c} \right)\left< a_{c} \right>-\mathrm{i}gN\left<
	\sigma_{i}^{-} \right>+\eta\label{eq:ac}\ ,\\ 
	\frac{\mathrm{d}}{\mathrm{d}t}\left< \sigma_{i}^{-}
	\right>&=-\left( \frac{\gamma_{\text{hom}}}{2}+\gamma_{\text{hom}}\bar{n}\left( T,\omega_{-}^{\text{I}} \right)+\gamma_{p}+
	\mathrm{i}\Delta_{-}^{\text{I}}\right)\left< \sigma_{i}^{-} \right>\notag\\
	&+\mathrm{i}g\left<
	\sigma_{i}^{z}a_{c}
	\right>\label{eq:sm}\ ,\\
	\frac{\mathrm{d}}{\mathrm{d}t}\left< \sigma_i^{z} \right>&=-\mathrm{i}2g\left( \left<
	\sigma_i^{+}a_{c}
	\right>-\left< \sigma_i^{-}a_{c}^{\dagger } \right> \right)-\gamma_{\text{hom}}\left( \left<
        \sigma_i^{z}
        \right> +1 \right)\notag\\
				&-2\gamma_{\text{hom}}\bar{n}\left( T,\omega_{-}^{\text{I}} \right)\left< \sigma_{i}^{z} \right>\ ,
	\label{eq:first_eqs}
\end{align}
which also includes equations for $\left< a_{c} \sigma_{i}^{+} \right>$, $\left< a_{c}
\sigma_{i}^{z} \right>$, $\left< \sigma_{i}^{+}\sigma_{j}^{-} \right>$, $\left<
a_{c}^{\dagger }a_{c} \right>$, $\left< a_{c}\sigma_{i}^{-} \right>$, $\left<
a_{c}^{\dagger }a_{c}^{\dagger } \right>$, $\left< \sigma_{i}^{-}\sigma_{j}^{-} \right>$,
$\left< \sigma_{i}^{z}\sigma_{j}^{+} \right>$, $\left< \sigma_{i}^{z}\sigma_{j}^{z}
\right>$ and $\left< a_{c}^{\dagger }a_{c} \sigma_{i}^{z}
\right>$. In order to truncate the system higher order terms in the equations are expanded in a
well defined way~\citep{kubo1962generalized}, then higher order cumulants are
neglected~\citep{meiser2009prospects,henschel2010cavity}. The equations again are written in a frame rotating
with the cavity probe frequency $\omega_{p}$.
Despite the inhomogeneous broadening, which is included by the nonradiative dephasing of
the spins, we assume that all spins are equal so that we only have to include the
equations for one spin $i$ and pairs $i,j$~\cite{meiser2009prospects}. This set of
equations can be integrated numerically to study the dynamics of the coupled system. We
note that when we fix $\left< \sigma_{i}^{z} \right>=-1$ we immediately arrive at the
model discussed in Sec.~\ref{subsec:HO} with $\gamma=\gamma_{\text{hom}}+2\gamma_{p}$.

In our experiment we measure the avoided crossing for different temperatures of the
environment and compare it to the results of our model. First we note that the steady
state of the inversion as a function of the temperature $\left< \sigma_{i}^{z} (T)
\right>_{\text{st}}$ can be written as~\citep{bushev2011ultralow}
\begin{align}
	\left< \sigma_{i}^{z} (T) \right>_{\text{st}}&=\frac{1}{1+2\bar{n}\left(
	T,\omega_{-}^{\text{I}} \right) }\left< \sigma_{i}^{z}(T=0) \right>_{\text{st}}\notag\\&=\tanh{\left(  \frac{\hbar\omega_{i}}{2k_{B}T}
	\right)}\left< \sigma_{i}^{z}(T=0) \right>_{\text{st}}\ .
	\label{eq:sz_st}
\end{align}
For higher temperatures $\left< \sigma_{i}^{z} (T) \right>_{\text{st}}$ is reduced and
therefore the effective number of NVs that take part in the dynamics. In the model
equations this is represented by the last term in Eq.\ref{eq:sm} for the
polarization involving  $\left< a_{c} \sigma_{i}^{z} \right>$, which leads to a cutoff for
the coupling at higher $T$. As a zero field approximation we thus write
\begin{align}
	\Omega(T)=g\sqrt{N\tanh{\left(  \frac{\hbar\omega^{\text{I}}_{-}}{2k_{B}T}\right)}}\ ,
	\label{eq:dT}
\end{align}
where we replaced $\omega_{i}$ by the center frequency $\omega^{\text{I}}_{-}$.
Equation~\eqref{eq:dT} should give an approximate description of the reduction of the Rabi
splitting with increasing temperature.

This treatment however neglects the
presence of the $m_{S}=+1$ state which will also be populated with increasing $T$.
Including this level and assuming $\bar{n}\left( T,\omega_{+}^{\text{I}}
\right)\approx\bar{n}\left( T,\omega_{-}^{\text{I}} \right)$ we find the population
difference between the $m_{S}=0$ and $m_{S}=-1$ state to be 
\begin{align}
	\frac{N_{-1}(T)}{N}-\frac{N_{0}(T)}{N}=-\frac{1}{1+3\bar{n}\left( T,\omega_{-}^{\text{I}}
	\right) }\ .
	\label{eq:3nbar}
\end{align}
This suggests
\begin{align}
	\tilde{\Omega}(T)=g\sqrt{ \frac{N}{1+3\bar{n}\left( T,\omega_{-}^{\text{I}}
	\right) }}
	\label{eq:dT2}
\end{align}
to be a better description for the temperature dependence of the coupling.

In a second step we integrate the whole hierarchic set of equations
numerically for $\omega_{c}=\omega_{-}^{\text{I}}$ and varying pump frequency
$\omega_{p}$.
From $\left| \left< a_{c} \right>_{\text{st}}\right|^{2} \left( \omega \right)$ we
determine the Rabi splitting for different temperatures from which we obtain the
collective coupling. This can be compared to our
measurements and the approximations in Eqs.~\ref{eq:dT} and ~\ref{eq:dT2}. 
\begin{figure}[tbp]
	\centering
	\includegraphics[width=0.45\textwidth]{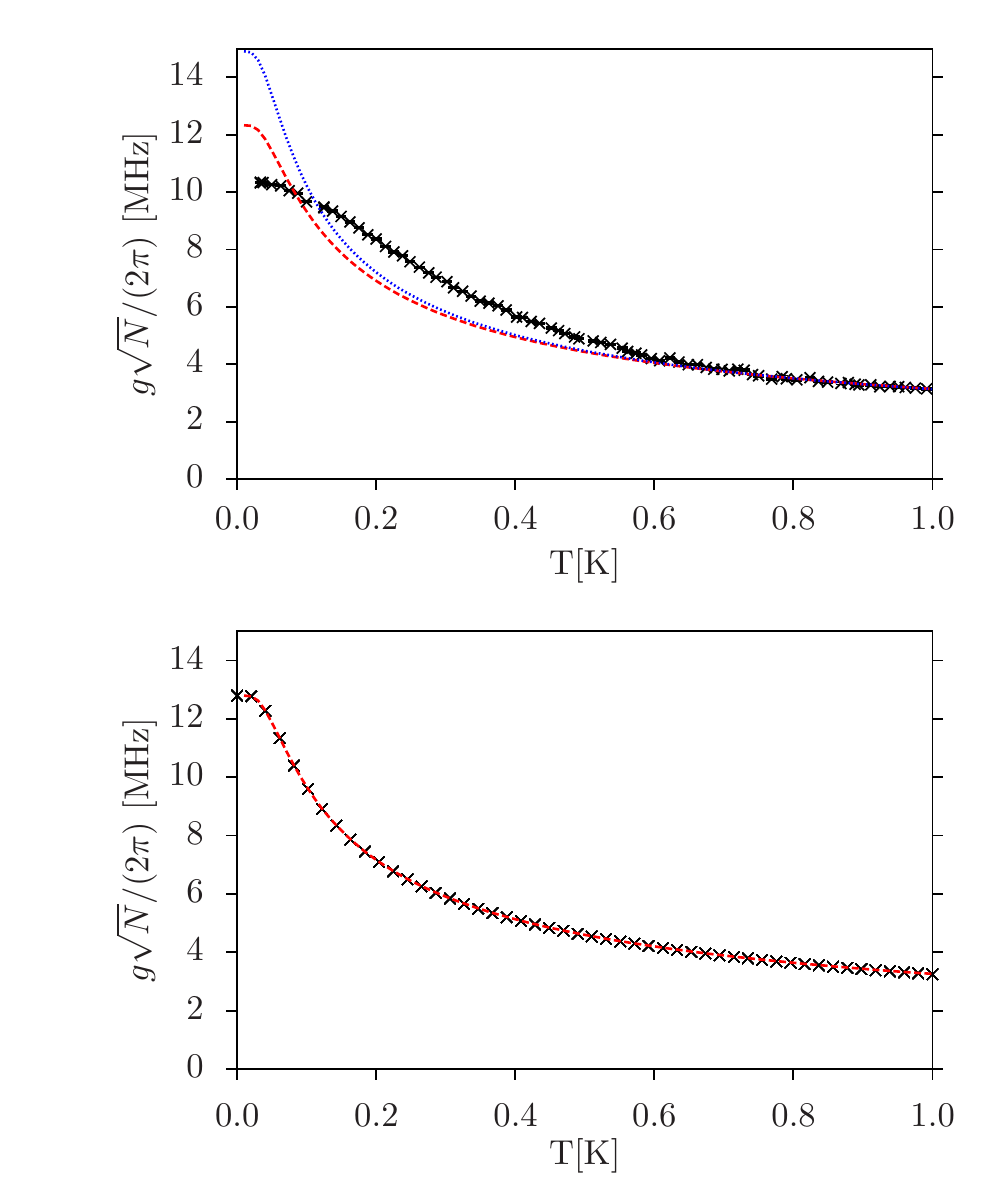}
	\caption{(Color online) (a) Collective coupling strength obtained from our
	measurements for different temperatures (black markers) and $\Omega(T)=a_{0}\left[\tanh{\left(
	\hbar\omega^{\text{I}}_{-}/(2k_{B}T)\right)}\right]^{1/2}$ (red dashed line), where
	$a_{0}$ was chosen to match the high temperature behavior. Including
	the $m_{S}=+1$ state via Eq.~\ref{eq:dT2} gives the result denoted by the blue dotted line. (b)
	Collective coupling strength determined from the results of the numerical integration of the
	coupled equations with $g\sqrt{N}=a_{0}$ ($a_{0}$ determined from (a), black
	markers) and $\Omega(T)$ (red dashed line).}
	\label{fig:coupling_vs_T_data_and_gamma_p}
\end{figure}

In Fig.~\ref{fig:coupling_vs_T_data_and_gamma_p}(a) we show that the measured coupling
strength is in disagreement with the Eq.~\ref{eq:dT}, which is most significant for very
low temperatures. Both functions Eq.~\ref{eq:dT} and Eq.~\ref{eq:dT2} fail to capture the
behavior for low temperatures. The disagreement is even more pronounced as we include the
effect of the $m_{S}=+1$ state. So far we found no definite explanation for the
disagreement. However, one possible explanation would be that for low temperatures not all
defects are ``active``. As temperature increases, more NVs become available but at the
same time the number of NVs taking part in the dynamics is proportional to
$\tanh{\left(\hbar\omega^{\text{I}}_{-}/(2k_{B}T)\right)}$.

The collective coupling strength determined from the numerical integration of the coupled
equations, shown in Fig.~\ref{fig:coupling_vs_T_data_and_gamma_p}(b), exactly follows
Eq.~\ref{eq:dT}. This shows that the assumption $\Omega(T)\propto 	\left< \sigma_{i}^{z}
(T) \right>_{\text{st}}$ is reasonable. However, the almost constant value of the
coupling strength $\Omega(T)$ found in the experimental data for very low $T$ cannot be
explained in the above theoretical model. As possible explanations one might think of a
reduced thermal excitation probability due to spin-spin coupling or an effective reduction
of active NV centers close to zero temperature. Interestingly the same behavior is also
found in measurements of dispersive shift of the cavity mode as a function of temperature
by the offresonant spin ensemble at zero magnetic field.


\subsection{Detailed modeling and reconstruction of inhomogeneous distributions}\label{subsec:resolvent}

The simplified model descriptions discussed above in Secs.~\ref{subsec:HO} and
~\ref{subsec:gamma_p} provided for an analytically tractable and qualitatively correct
description of the effect of an inhomogeneous broadening of the ensemble. This also
allows to get a fairly good estimate for the total width of the frequency distribution of
the ensemble. However, such an effective width model inherently is connected to the
assumption of a Lorentzian shape of the ensemble frequency distribution. In actual crystals
such an assumption is not obvious and other distributions of local field variations and
strain distributions are possible as well. 

To obtain more accurate information about the distribution we will now use an improved
model based on the resolvent formalism to treat the coupling between a central oscillator
(the mode) and the spin degrees of
freedom~\citep{kurucz2011spectroscopic,cohen1992atom}. Here each frequency class of
spins is treated individually. For low temperatures virtually all spins  are in the
$m_{S}=0$
state, i.e. the lower state of our effective two-level system and their excitation
properties can be approximated by a frequency distributed set of oscillators
(Holstein-Primakoff approximation).
 
We define creation and annihilation operators for the  corresponding ensemble oscillators
representing a subclass of two-level systems with equal frequency via
\begin{align}
	\sigma^{z}_{j}=-1+2a_{j}^{\dagger}a_{j}\quad \text{and} \quad
	\sigma_{j}^{+}=a_{j}^{\dagger}\sqrt{1-a_{j}^{\dagger}a_{j}}\approx a_{j}^{\dagger}\ .
	\label{eq:HP_trafo}
\end{align}
The approximation in Eq.~\ref{eq:HP_trafo} is justified as long as the number of
excitations in each ensemble is much smaller than the number of spins in this energy
region. In our experimental setup this is very well justified and we thus obtain the
unperturbed part of the Hamiltonian as
\begin{align}
	\operatorname{H}_{0}^{\prime}&=\omega_{c}a_{c}^{\dagger}a_{c}+\sum^{N}_{j}\omega_{j}a_{j}^{\dagger}a_{j}\ ,
	\label{eq:Hamiltonian_HP}
\end{align}
and the interaction term
\begin{align}
	\operatorname{V}^{\prime}=\sum^{N}_{j}\left(g_{j}a_{j}^{\dagger}a_{c}+g^{*}_{j}a_{j}a^{\dagger}_{c}\right)\
	,
	\label{eq:Interaction_HP}
\end{align}
which constitute
$\operatorname{H}^{\prime}=\operatorname{H}_{0}^{\prime}+\operatorname{V}^{\prime}$. 
In this section we account for the decay of cavity excitations and the spontaneous decay of the spins by
introducing nonzero imaginary parts of the corresponding transition frequencies
$\operatorname{Im}\omega_{c}=-\kappa$ and
$\operatorname{Im}\omega_{j}=-\frac{1}{2}\gamma_{\text{hom}}$ and
the resolvent of the Hamiltonian $\operatorname{H}^{\prime}$ is defined as
$\operatorname{G}\left( z \right)=1/\left( z-\operatorname{H}^{\prime}\right)$.

Let us consider the state $\ket{\varphi_{c}}=\ket{1_{c},0,\dots,0}$ where we have one
photon in the cavity and no excitation in the ensemble.

The matrix element of the resolvent $\operatorname{G}_{cc}\left( z
\right)=\bra{\varphi_{c}}\operatorname{G}\left( z \right)\ket{\varphi_{c}}$ can be written
as
\begin{align}
	G_{cc}\left( z \right)=\frac{1}{z-E_{c}-R_{cc}\left( z \right)}\ ,
	\label{eq:elem_resolv}
\end{align}
where we define the matrix element of the level shift operator
\begin{align}
	R_{cc}\left( z \right)&=V^{\prime}_{cc}+\sum_{i\neq c}V^{\prime}_{ci}\frac{1}{z-E_{i}}V^{\prime}_{ic}\notag\\
	&+\sum_{i\neq
c}\sum_{j\neq c}
V^{\prime}_{ci}\frac{1}{z-E_{i}}V^{\prime}_{ij}\frac{1}{z-E_{j}}V^{\prime}_{jc}+\dots \ ,
	\label{eq:elem_levelshift}
\end{align}
where $V^{\prime}_{kl}=\bra{\varphi_{k}}\operatorname{V}^{\prime}\ket{\varphi_{l}}$ and
$E_{c}=\bra{\varphi_{c}}\operatorname{H}^{\prime}_{0}\ket{\varphi_{c}}$. States
$\ket{\varphi_{i}}$ with $i\neq c$ are states where the excitation is absorbed in the
ensemble spin $i$. We note that $V^{\prime}_{cc}=0$ and that $V^{\prime}_{ij}=0$ for
$i,j \neq c$ since our Hamiltonian does not include spin-spin interaction. Only the second
term in Eq.~\ref{eq:elem_levelshift} remains and we can write
\begin{align}
	R_{cc}\left( z \right)=\sum_{i\neq c}\frac{\left| g_{i}\right|^{2} }{z-E_{i}}\ .
	\label{eq:elem_levelshift_reduced}
\end{align}
Introducing the coupling density profile $\rho\left( \omega \right)\equiv\sum_{j}\left|
g_{j}\right|^{2}\delta\left( \omega-\operatorname{Re}\omega_{j} \right) $ as it is done
in~\cite{kurucz2011spectroscopic} leads to
\begin{align}
	R_{cc}\left( z \right)=\int \frac{\rho\left( \omega
	\right)}{z-\omega+\frac{\mathrm{i}}{2}\gamma_{\text{hom}}}\mathrm{d}\omega\ .
	\label{eq:elem_levelshift_continous}
\end{align}
Approaching the branch cut at $\omega-\frac{\mathrm{i}}{2}\gamma_{\text{hom}}$ we write
\begin{align}
	\lim_{\eta\rightarrow 0^{+}} R_{cc}\left( \omega-\frac{\mathrm{i}}{2}\gamma_{\text{hom}} \pm \mathrm{i}\eta\right)=R^{\pm}_{cc}\left(
	\omega-\frac{\mathrm{i}}{2}\gamma_{\text{hom}}\right)\ .
	\label{eq:elem_levelshift_atcut}
\end{align}
Using $\lim_{\eta\rightarrow 0^{+}} \frac{1}{x\pm
\mathrm{i}\eta}=\mathcal{P}\frac{1}{x}\mp \mathrm{i}\pi \delta\left( x \right)$, where
$\mathcal{P}$ denotes the Cauchy principal value, we find
\begin{align}
	R^{\pm}_{cc}\left( \omega-\frac{\mathrm{i}}{2}\gamma_{\text{hom}}
	\right)=\mathcal{P}\int \frac{\rho\left( \omega^{\prime}
	\right)}{\omega-\omega^{\prime}}\mathrm{d}\omega^{\prime}\mp \mathrm{i}\pi\rho\left(
	\omega \right)\ .
	\label{eq:elem_levelshift_real_im}
\end{align}

Experimentally we probe the transmission of a weak probe signal amplitude through the
cavity as a function of frequency. The position and shape of the weak field transmission
resonances can be determined from the complex poles of $G^{+}_{cc}\left( \omega
\right)=1/\left( \omega-\omega_{c}-R^{+}_{cc}\left( \omega \right) \right)$, which
contains the spin energy distribution on the right hand side. We can therefore extract
information about the coupling density $\rho\left( \omega
\right)=-\frac{1}{\pi}\operatorname{Im}R^{+}_{cc}\left(
\omega-\frac{\mathrm{i}}{2}\gamma_{\text{hom}} \right)$ by carefully analyzing the
measured transmission spectrum. For small $\gamma_{\text{hom}}$ the reconstruction
simplifies to:
\begin{align}
	\rho\left( \omega \right)\approx -\frac{1}{\pi}\operatorname{Im}R^{+}_{cc}\left( \omega
	\right)+\frac{1}{2\pi}\frac{\partial \operatorname{Re}R^{+}_{cc}\left( \omega
	\right)}{\partial \omega}\gamma_{\text{hom}}\ ,
	\label{eq:coupling_dens_approx}
\end{align}
where $\operatorname{Im}R_{cc}^{+}\left( \omega-\frac{\mathrm{i}}{2}\gamma_{\text{hom}}
\right)$ is expanded in a Taylor series around $\gamma_{\text{hom}}=0$. 
We can therefore directly use the frequency distribution of the transmitted signal
 $\left| \left< a_{c} \right>_{\text{st}}\right| ^{2}$ via
\begin{align}
	\left| \left< a_{c}
	\right>_{\text{st}}\right|^{2}&\propto \left| G^{+}_{cc}\left( \omega,\omega_{c} \right)\right| ^{2}\notag\\
	&= \frac{1}{\left( \omega-\omega_{c}-\operatorname{Re}R^{+}_{cc}\left( \omega
	\right) \right)^{2}+\left( \kappa+ \operatorname{Im}R^{+}_{cc}\left( \omega
	\right) \right)^{2}}
	\label{eq:lorentz}
\end{align}
to determine $R^{+}_{cc}\left( \omega \right)$. Let us point out here that
Eq.~\ref{eq:lorentz} exhibits a Lorentzian shape as a function of $\omega_{c}$ with peak
position $\omega-\operatorname{Re}R^{+}_{cc}\left( \omega \right)$ and peak width of
$2\left(\kappa+\operatorname{Im}R^{+}_{cc}\left( \omega \right)\right)$.  After
extraction of the relevant parameters which determine  $R^{+}_{cc}\left( \omega \right)$
from the measured spectra, we can simply use Eq.~\ref{eq:coupling_dens_approx} to find the
coupling density $\rho\left( \omega \right)$. Assuming that all spins are coupled with
equal strength, one finds that $\rho\left( \omega \right)=g^{2}N\mathcal{M}\left( \omega
\right)$ where $\mathcal{M}\left( \omega \right)$ is the frequency distribution of the
spins.

\begin{figure}[t]
	\centering
	\includegraphics[width=0.5\textwidth]{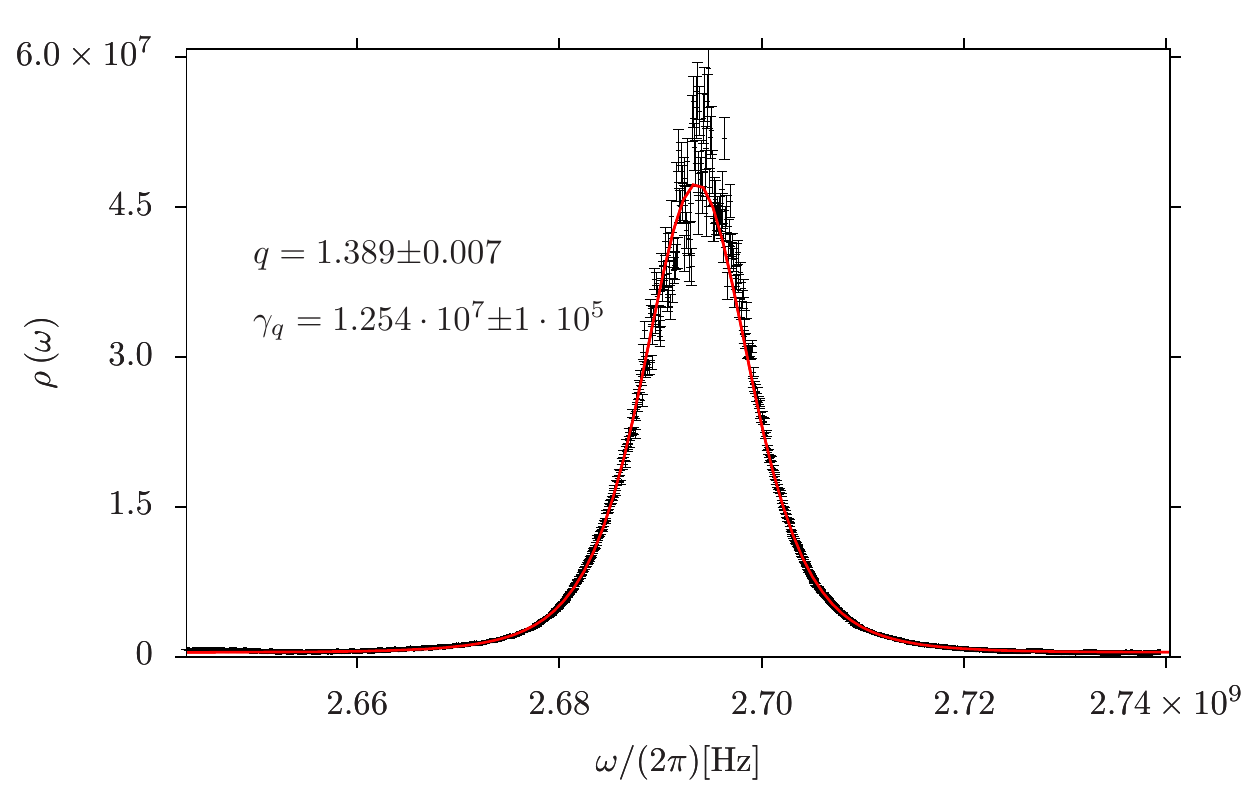}
	\caption{(Color online) Coupling density of the ensemble determined from an exemplary
	transmission measurement at $T=\unit{20}{\milli\kelvin}$. The errorbars are calculated from the
	uncertainties in $R_{cc}^{+}\left( \omega \right)$ which we obtain from fitting
	Eq.~\ref{eq:lorentz} to the data. We fit the q-Gaussian (red) in Eq.~\ref{eq:QGauss} to the
	coupling density to determine the width and the behavior of the tails of the
	distribution.}
	\label{fig:coupling_density_Qgauss}
\end{figure}

The shape of the coupling density, i.e. the frequency distribution of the spins,  plays an
important role in the cavity ensemble
interaction~\cite{kurucz2011spectroscopic,diniz2011strongly}. If the coupling density
falls off sufficiently fast with distance from the center, the width of the Rabi peaks
will decrease with increasing collective coupling strength $g\sqrt{N}$. For the spin
frequency distribution, the limiting case is the Lorentzian coupling density profile, for
which the width of the Rabi peaks is independent of $g\sqrt{N}$~\citep{diniz2011strongly}.
Any distribution falling of faster than $1/\omega^{2}$ will provide a decrease of the
width of the Rabi peaks. Moreover, knowing the coupling density gives us the opportunity
to study the transmission through the cavity for different parameter ranges via
Eq.~\ref{eq:elem_resolv}.

To determine $\rho\left( \omega \right)$ the raw data have to be rearranged, since in the experiment we cannot simply vary
$\omega_{c}$ but we shift the center frequency of the spins by a magnetic field. We
therefore shift each scan by $\omega_{c}-\omega_{-}^{\text{I}}\left( B \right)$ to obtain
fixed ensemble frequencies and tuning of the cavity frequency. As the only significant
quantity is the detuning between the cavity and the ensemble this does not change the
dynamics. For fixed $\omega$ we fit a Lorentzian to $\left| G^{+}_{cc}\left(
\omega,\omega_{c} \right)\right| ^{2}$ to determine $R^{+}_{cc}\left( \omega \right)$ in
order to calculate $\rho\left( \omega \right)$. We plot an exemplary result for
$\rho\left( \omega \right)$ in Fig.~\ref{fig:coupling_density_Qgauss}. To determine the
behavior of the tails of the distribution we fit the function
\begin{align}
	L\left( \omega \right)=b+I\left[ 1-\left( 1-q \right)\frac{\left(
	\omega-\omega_{0} \right)^{2}}{a} \right]^{\frac{1}{1-q}}\ ,
	\label{eq:QGauss}
\end{align}
to the data. $L\left( \omega \right)$ is related to the q-Gaussian, a Tsallis
distribution. The dimensionless parameter $1<q<3$ determines how fast the tails of the
distribution fall off,
while $a$ is related to the width. The actual width (FWHM) is given by $\gamma_{q}=2
\sqrt{\frac{a\left( 2^{q}-2 \right)}{2q-2}}$. For $q\rightarrow 1$ we recover a Gaussian
distribution, while for $q=2$ we find a Lorentzian distribution. The wings fall of as
$ 1/\omega^{2/(q-1)}$.

From the fit in Fig.~\ref{fig:coupling_density_Qgauss} we find values for $q=1.389 \pm
0.007$ and $\gamma_{q}/(2\pi)=\unit{(12.54 \pm 0.10)}{\mega\hertz}$. The temperature
during the measurement was $\unit{20}{\milli\kelvin}$.

\begin{figure}[t]
	\centering
	\includegraphics[width=0.48\textwidth]{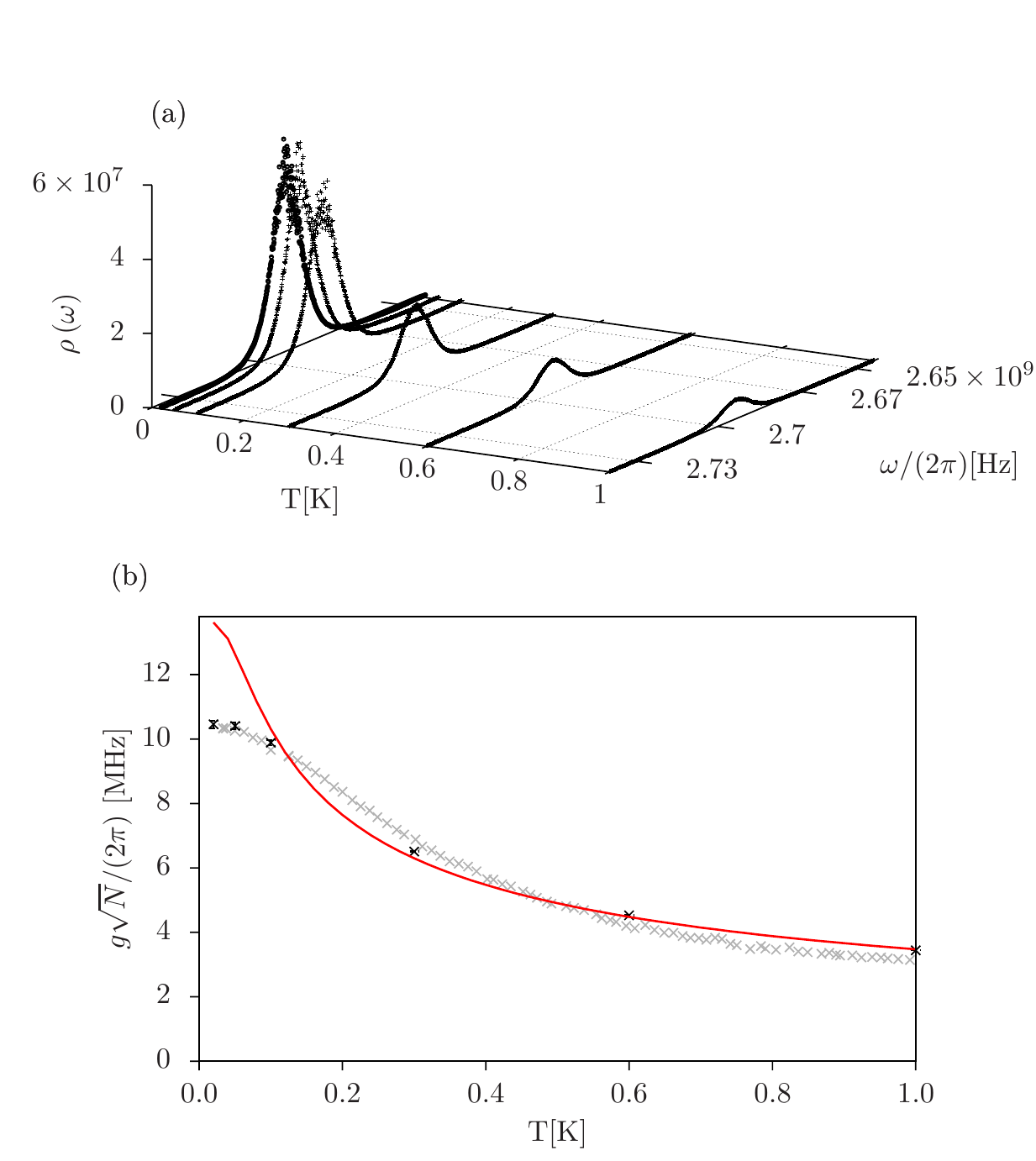}
	\caption{(Color online) In (a) we show the coupling density for different temperatures. The $q$-parameter and width $\gamma_{q}$ of the coupling density do not change
	significantly with increasing temperature. The coupling
	$g\sqrt{N}=\left(\int \rho\left( \omega \right)\mathrm{d}\omega\right)^{1/2}$
	is reduced, as can be seen in (b). The red line shows $\Omega(T)=a_{0}\left[\tanh{\left(
	\hbar\omega^{\text{I}}_{-}/(2k_{B}T)\right)}\right]^{1/2}$ with $a_{0}$ chosen to match
	the high temperature limit of the data (black markers). To compare $\Omega(T)$ with our previous results
	determined directly from the splitting we plot the data already shown in
	Fig.~\ref{fig:coupling_vs_T_data_and_gamma_p} using grey markers.}
	\label{fig:coupling_density_fence_plot}
\end{figure}

The same analysis is performed for data measured for $T=\unit{20 - 1000}{\milli\kelvin}$,
see Fig~\ref{fig:coupling_density_fence_plot}(a). The fitted curves are not shown.
For the resulting coupling densities we find almost no change in the width or the
q-parameter.
However, with increasing temperature $\int \rho\left( \omega
\right)\mathrm{d}\omega=g^{2}N$ is reduced, as
can be seen in Fig.~\ref{fig:coupling_density_fence_plot}(b). The resulting coupling
strength is in agreement with the results obtained from the analysis in
Sec.\ref{subsec:gamma_p}. It also reproduces the unexpected behavior for small values of
$T$. 

\begin{figure}[t]
	\centering
	\includegraphics[width=0.45\textwidth]{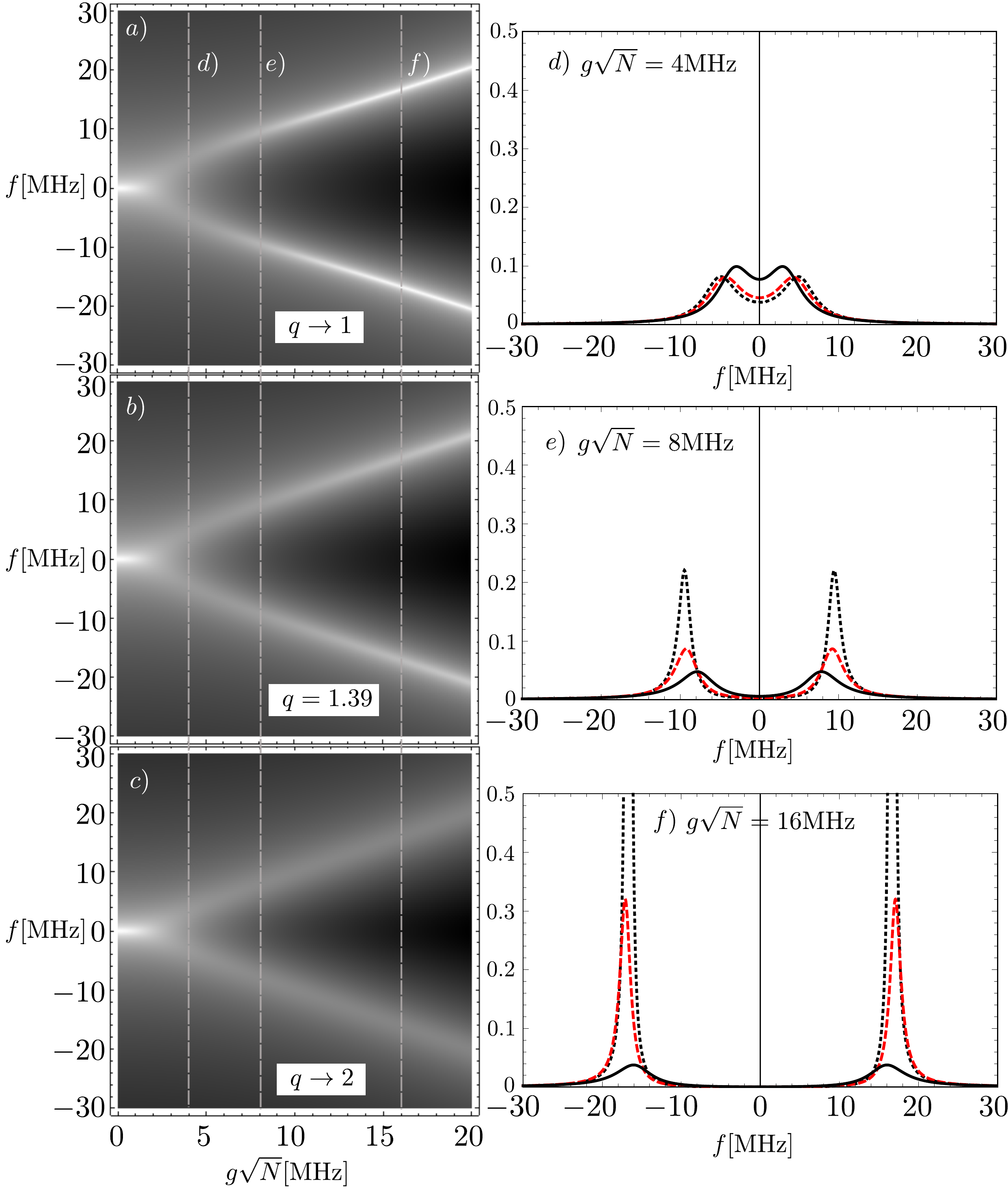}
	\caption{(Color online) We plot $\left|G_{cc}\left( \omega \right) \right| ^{2}$ as a
	function of the collective coupling strength $g\sqrt{N}$ and for three different values
	of the q-Gaussian parameter $q$ (grayscale figures on the left are plotted on a
	logarithmic scale). For $g\sqrt{N}=\unit{4,8\, \text{and}\, 16}{\mega\hertz}$, marked
	by $\left.\text{d}\right)$, $\left.\text{e}\right)$ and $\left.\text{f}\right)$ we show the transmission for all three values of $q$ together. The
	transmission for the Gaussian coupling density with $q\rightarrow 1$ (dotted black
	line), for the coupling density of our ensemble with $q=1.39$ (dashed red line) and for the Lorentzian coupling
	density with $q\rightarrow 2$ (solid black line) are shown on the right. The parameters were chosen to
	$\gamma_{q}/(2\pi)=\unit{10}{\mega\hertz}$, $\kappa/(2\pi)=\unit{0.4}{\mega\hertz}$ and
	$\gamma_{\text{hom}}/(2\pi)=\unit{1}{\hertz}$.}
	\label{fig:transmission_vs_g_QGAUSS}
\end{figure}

\begin{figure}[tbp]
	\centering
	\includegraphics[width=0.45\textwidth]{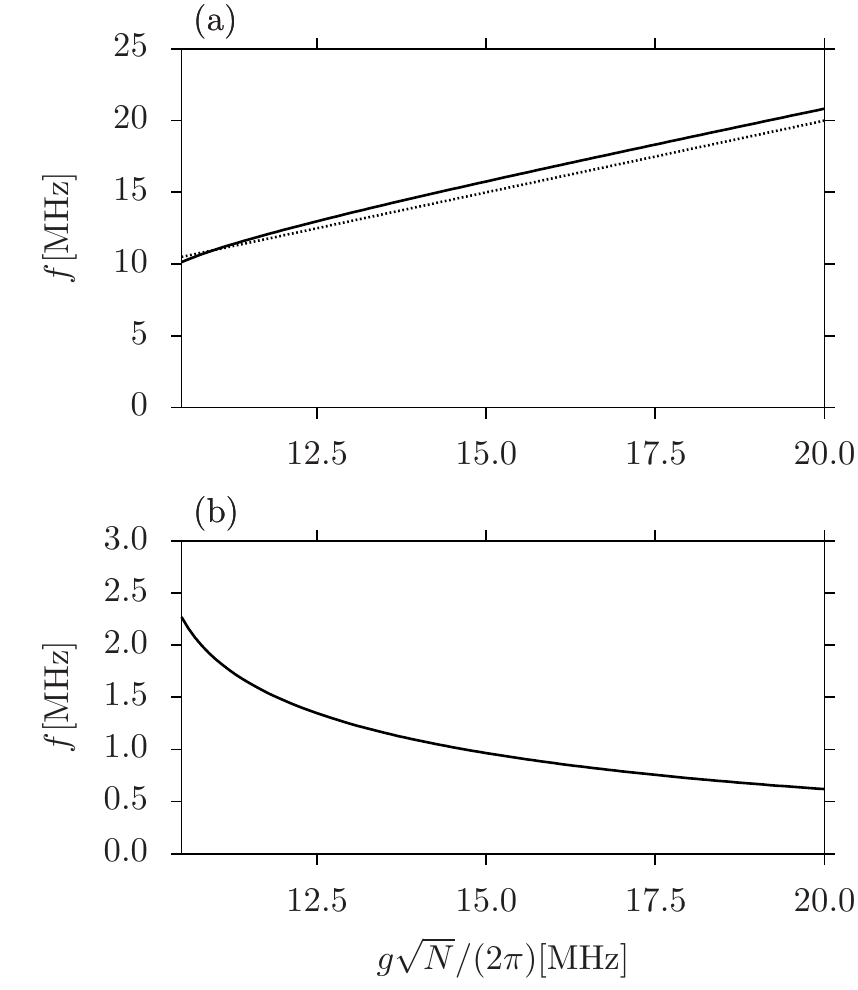}
	\caption{(a) Real part and (b) modulus of the imaginary part of
one of the complex poles of $G^{+}_{cc}\left( \omega \right)$ in the regime where
$g\sqrt{N}/(2\pi)> \gamma_{q}/(2\pi)=\unit{10}{\mega}\hertz$. The collective coupling
strength $g\sqrt{N}$ is shown as dotted line in (a). With increasing $g\sqrt{N}$ the modulus of
the imaginary part, proportional to the width of the Rabi
peaks, is reduced.}
	\label{fig:width_position_vs_g_QGAUSS_q_1_39}
\end{figure}

To study the behavior of an ensemble following a q-Gaussian distribution with the
q-parameter we determined, we show $\left|G_{cc}\left( \omega \right) \right| ^{2}$ for
$\omega_{c}=\omega_{0}$ as a function of the collective coupling strength $g\sqrt{N}$ in
Fig.~\ref{fig:transmission_vs_g_QGAUSS} $\left.\text{b}\right)$. The transmission for an
ensemble with Gaussian distribution ($q\rightarrow 1$) is shown in
Fig.~\ref{fig:transmission_vs_g_QGAUSS} $\left.\text{a}\right)$ and for a Lorentzian
distribution ($q\rightarrow 2$) in Fig.~\ref{fig:transmission_vs_g_QGAUSS}
$\left.\text{c}\right)$. The width parameter $a$ is chosen accordingly to ensure that
$\gamma_{q}$ is the same for all three cases. In Figs.~\ref{fig:transmission_vs_g_QGAUSS}
$\left.\text{d}\right)$-$\left.\text{f}\right)$ the transmission for the three ensemble
types is shown for $g\sqrt{N}=\unit{4,8\, \text{and}\, 16}{\mega\hertz}$, respectively.
For the ensemble with Lorentzian distribution (solid black line) we see that the width of
the Rabi resonances remains constant with increasing collective coupling. For the
Gaussian distribution (dotted black line) and the intermediate distribution with $q=1.39$
(dashed red line) we find a decrease in the peak width for increasing collective coupling.
In Fig.~\ref{fig:width_position_vs_g_QGAUSS_q_1_39} we plot the real and imaginary part of
one of the complex poles of $G^{+}_{cc}\left( \omega \right)$, determining position and
width of one of the Rabi peaks. As we focus on the resonant case the spectrum is
symmetric. We chose $\gamma_{\text{hom}}=0$, $\kappa/(2\pi)=\unit{0.4}{\mega\hertz}$,
$\gamma_{q}/(2\pi)=\unit{10}{\mega\hertz}$ and $q=1.39$. This again shows the
decreasing width of the Rabi Peaks as $g\sqrt{N}$ increases.

We therefore assume that for our ensemble it is possible to increase the lifetime of the
collective states by increasing the collective coupling. This could be achieved by a
further decrease of the mode volume or an increase of the NV density in the sample.

\begin{figure}[tbp]
	\centering
	\includegraphics[width=0.4\textwidth]{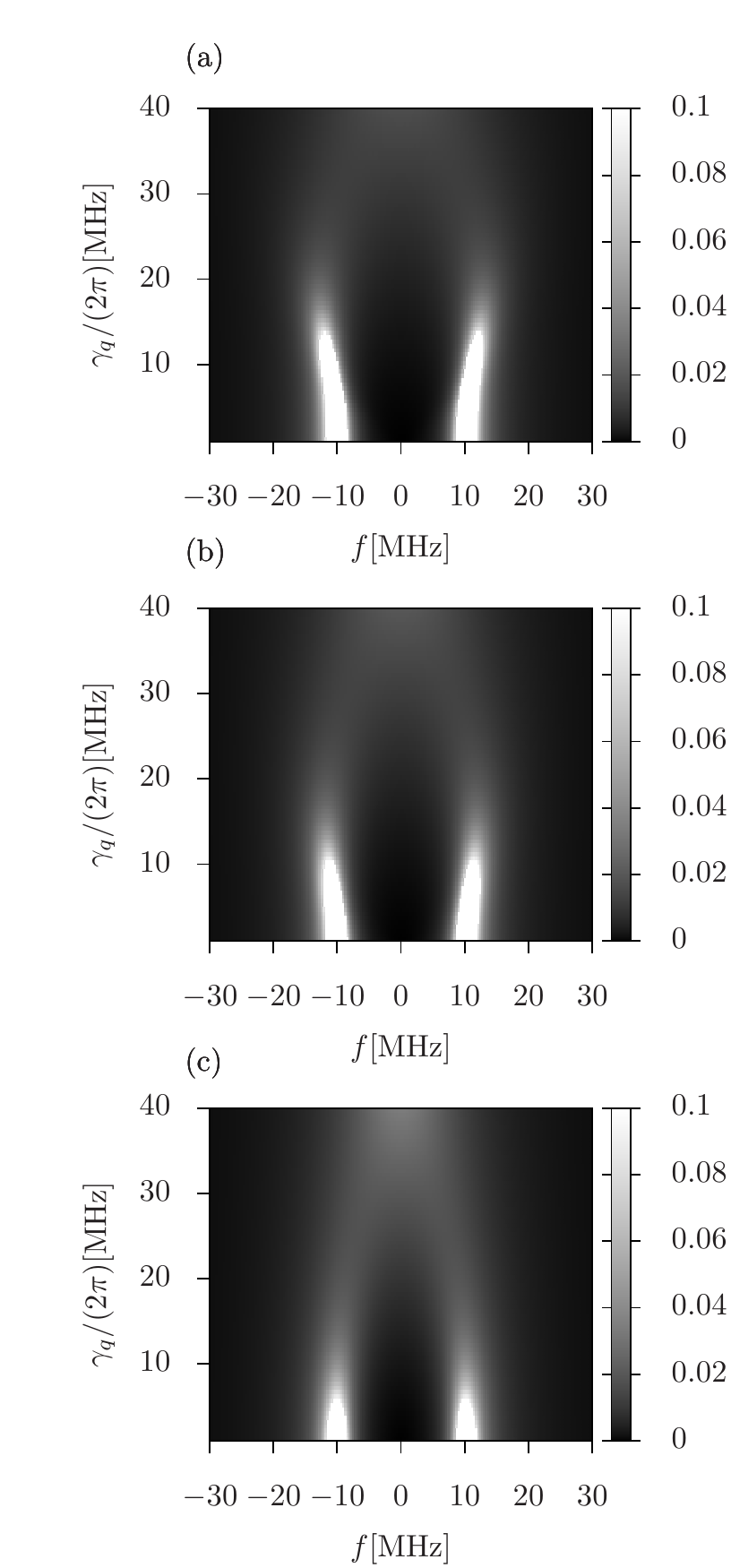}
	\caption{$\left|G_{cc}\left( \omega \right) \right| ^{2}$ on resonance as a
	function of the inhomogeneous width $\gamma_{q}$ for an ensemble with (a) Gaussian coupling
density, (b) q-Gaussian $(q=1.39)$ coupling density and (c) Lorentzian coupling density. To keep the transmission visible for
	large values of $\gamma_{q}$ the range of the color coding is
	limited to $[0,0.1]$. The parameters were chosen to
	$g\sqrt{N}/(2\pi)=\unit{10}{\mega\hertz}$, $\kappa/(2\pi)=\unit{0.4}{\mega\hertz}$ and
	$\gamma_{\text{hom}}/(2\pi)=\unit{1}{\hertz}$.}
	\label{fig:transmission_vs_gamma_q_QGAUSS_q_1_39}
\end{figure}

In Fig.~\ref{fig:transmission_vs_gamma_q_QGAUSS_q_1_39} we show the behavior of the
transmission with increasing $\gamma_{q}$. For the Lorentzian coupling density the
splitting of the resonance peaks is always reduced if $\gamma_{q}>0$. For coupling
densities falling off faster than $1/\omega^{2}$, as it is the case in (a) and (b), the
splitting of the resonance peaks is even slightly increased for $\gamma_{q}>0$, until the
peaks finally merge. Hence a finite but small enough inhomogeneous width might even mimic somewhat higher 
active spin numbers.

As bottom line we see that common coupling to a cavity mode can suppress dephasing of the
polarization, if the effective Rabi frequency is larger than the inhomogeneous width. This
could be interpreted as the effect that the exchange of the excitation between the
ensemble and the cavity is so fast, that there is no time for decoherence in the ensemble.


\section{A transmission line micro-maser with an inhomogeneous NV ensemble}\label{sec:maser}

A recent proposal to construct a laser operating on an ultra-narrow atomic clock transition
predicted very narrow optical emission above threshold ~\citep{meiser2009prospects}. Similar
ideas, employing the collective coupling between a cold atomic ensemble and a microwave cavity, have been
proposed to construct stable stripline oscillators in the microwave regime~\citep{henschel2010cavity}.

Here
we study the prospects of implementing such an oscillator by coupling a diamond to the CPW resonator.
At first sight in view of the MHz scale inhomogeneous broadening, one would expect fast dephasing. However,
as we have seen above, for strong enough coupling one observes a continuous rephasing of the polarization
inducing a long lived polarization and coherent Rabi oscillations. Thus one could expect
narrow microwave emission nevertheless.

Let us consider the case of the cavity mode tuned to resonance with the spin transition
$\ket{0}_{\text{I}}\rightarrow \ket{-}_{\text{I}}$, which is partially inverted by an
external incoherent pump. Such a pump could in principle be facilitated by optical pumping
and it can be consistently modeled by a reversed spontaneous decay. Alternatively one
could think of pulsed inversion by tailored microwave pulses or a time switching of the
magnetic bias field. 

Mathematically such incoherent pumping can be modelled by adding the terms
$-\frac{w}{2}\sum_{j=1}^{N}\left(\sigma_j^{-}\sigma_j^{+}\rho +\rho
\sigma_j^{-}\sigma_j^{+}-2\sigma_j^{+}\rho \sigma_j^{-} \right)$, where $w$ denotes the
pump rate, to the Liouvillian in Eq.~\ref{eq:Liou}.

\begin{figure}[h!]
	\centering
	\includegraphics[width=0.45\textwidth]{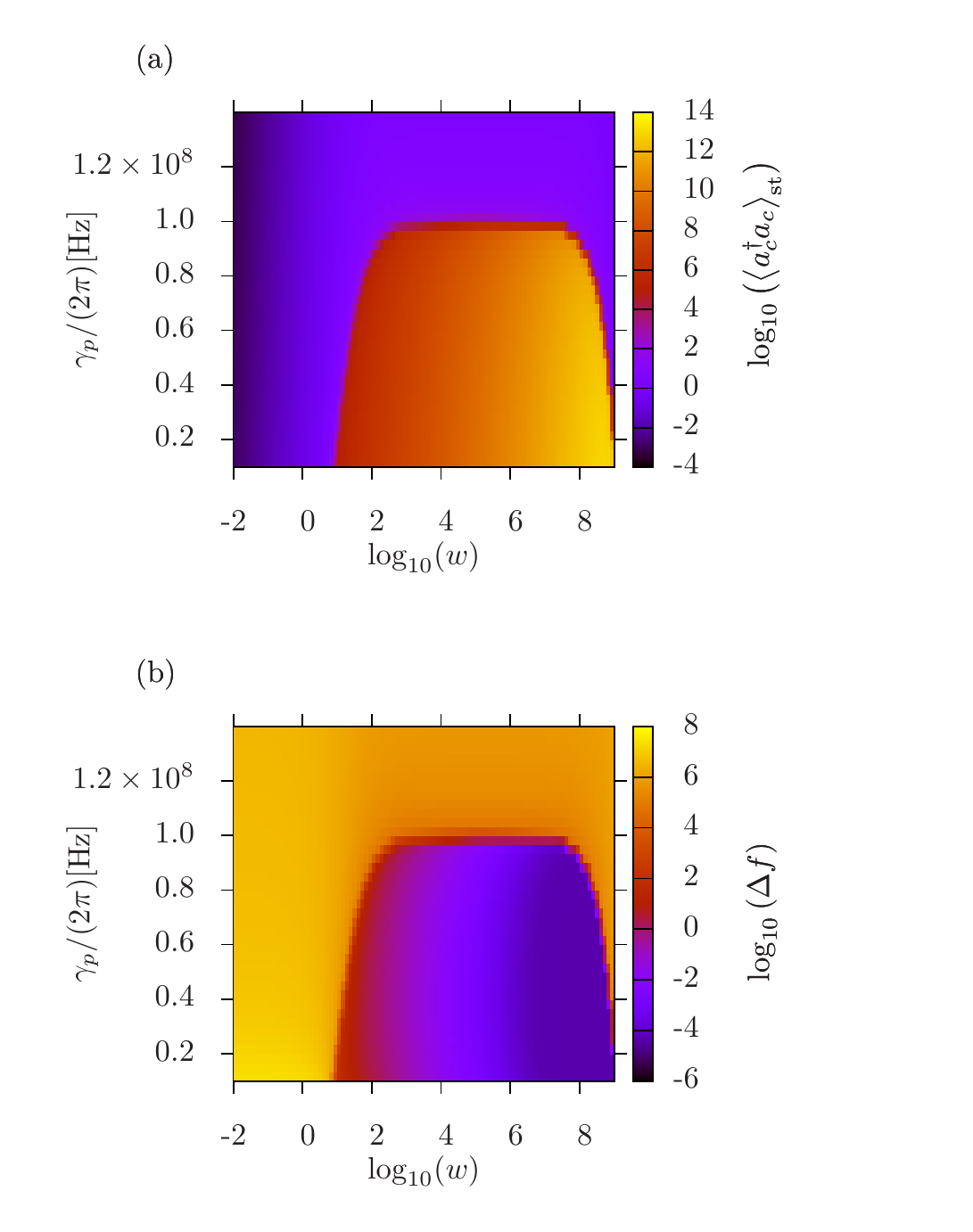}
	\caption{(Color online) In (a) we show the steady state number of photons in the cavity
	$\left< a_{c}^{\dagger}a_{c} \right>_{\text{st}}$ as a function of the incoherent pump
	rate $w$ and the inhomogeneous width $\gamma_{p}$. The resulting linewidth $\Delta f$ is
	shown in (b). For small $\gamma_{p}$ both, $\Delta f$ and $\left< a_{c}^{\dagger}a_{c} \right>_{\text{st}}$
	show rapid changes as $w$ becomes larger than $\gamma_{\text{hom}}$. The critical width of the
inhomogeneity is given by $\gamma_{p,\text{crit}}=g^{2}N/\kappa$. The parameters were
chosen to $N=10^{12}$, $g/(2\pi)=\unit{10}{\hertz}$,
$\kappa/(2\pi)=\unit{1}{\mega\hertz}$ and $\gamma_{\text{hom}}/(2\pi)=\unit{1}{\hertz}$.}
	\label{fig:laser1}
\end{figure}

For the explicit calculations, here we use the effective linewidth model, as outlined in Sec.~\ref{subsec:gamma_p} 
without any coherent pump. Hence the total phase symmetry of the system is
not broken and we assume $\left< a_{c} \right>=\left< a^{\dagger}_{c} \right>=\left<
\sigma^{\pm}_{j} \right>=0$. Starting from the master equation in Eq.~\ref{eq:Master} we
derive four coupled equations for $\left< \sigma^{z}_{j} \right>$, $\left<
a_{c}^{\dagger}a_{c} \right>$, $\left< a_{c}\sigma_{i}^{+} \right>$ and $\left<
\sigma_{i}^{+} \sigma_{j}^{-} \right>$. We used the cumulant expansion $\left<
a_{c}^{\dagger}a_{c} \sigma_{i}^{z} \right>=\left< a_{c}^{\dagger}a_{c} \sigma_{i}^{z}
\right>_{\text{cum}}+\left< a_{c}^{\dagger}a_{c} \right>\left< \sigma_{i}^{z} \right>$
which takes this simple form because of the total phase invariance. Assuming that the
higher order cumulant $\left< a_{c}^{\dagger}a_{c} \sigma_{i}^{z} \right>_{\text{cum}}$
can be neglected, we arrive at a closed set of four equations that can be solved
analytically.
To study the spectrum of the emitted light we calculate the two-time correlation function
$\left< a_{c}^{\dagger}\left( \tau \right)a_{c}\left( 0\right) \right>$ via the
quantum regression theorem. We switch to a frame rotating with $\omega_{c}$ and define
$\Delta=\omega_{-}^{\text{I}}-\omega_{c}$.
We obtain
\begin{widetext}
	\begin{align}
		\frac{\mathrm{d}}{\mathrm{d}\tau}\begin{pmatrix} \left< a_{c}^{\dagger}\left( \tau
			\right)a_{c}\left( 0\right) \right>  \\ \left< \sigma_{i}^{+}\left( \tau
			\right)a_{c}\left( 0\right) \right>  \end{pmatrix}=
			\begin{pmatrix} -\kappa & \mathrm{i} g N \\ -\mathrm{i} g \left<
				\sigma_{i}^{z} \right>_{\text{st}} & -\left(
				\frac{w+\gamma_{\text{hom}}}{2}+\gamma_{\text{hom}}\bar{n}\left( T,\omega_{-}^{\text{I}} \right)+\gamma_{p}-\mathrm{i}\Delta \right) \end{pmatrix}
		\begin{pmatrix} \left< a_{c}^{\dagger}\left( \tau
			\right)a_{c}\left( 0\right) \right>  \\ \left< \sigma_{i}^{+}\left( \tau
			\right)a_{c}\left( 0\right) \right>  \end{pmatrix}
		\label{eq:tt_corr_eqs}
	\end{align}
\end{widetext}
From Eq.~\ref{eq:tt_corr_eqs} we can calculate the spectrum via Laplace
transformation~\citep{meystre2007elements}. We show the linewidth of the obtained
spectrum in the resonant case ($\Delta=0$) for different values of the pump $w$ and the
inhomogeneous width $\gamma_{p}$, see Fig.~\ref{fig:laser1}.
The minimum linewidth is $\Delta f \approx
g^{2}/\kappa$. For small $\gamma_{p}$ the region where we find narrow linewidth emission
is characterized by $\gamma_{\text{hom}}<w<2 g^{2}N/\kappa$. The critical width of the
inhomogeneity is given by $\gamma_{p,\text{crit}}=g^{2}N/\kappa$.
Hence once we achieve enough pumping and coupling strength the system could provide for an extremely stable microwave
oscillator. 

\section{Conclusions}
We showed theoretically and experimentally that an ensemble of spins with an inhomogeneous
frequency distribution coupled to a cavity mode can exhibit strong coupling, where the
coherent energy exchange between mode and ensemble dominates cavity decay and polarization
dephasing. A detailed theoretical modeling connecting probe transmission and frequency
distribution allows to extract not only the effective coupling strength and particle
number but also the detailed frequency distribution. Interestingly for our dense NV
ensemble, the frequency distribution of the spins can be well described as a q-Gaussian
with $q=1.39$, showing that the wings of the distribution fall off faster than
$1/\omega^{2}$. The temperature dependence of the effective available spin number fits
quite well with expectations, except for an unexpected decrease for very low temperature.
In summary such an NV ensemble cavity QED system exhibits a prolonged lifetime of
the eigenstates of the coupled cavity-ensemble system~\citep{kurucz2011spectroscopic} and
has great potential as quantum interface between superconducting and optical qbits. The
long effective $T_1$ time could also be the basis of building a compact ultrastable
microwave oscillator if the strong coupling overcomes the dephasing from the inhomogeneous
broadening.

\section{Acknowledgment}
K.S. was supported by the DOC-fFORTE doctoral program. R.A. and T.N. were supported by CoQuS, C.K. by
FunMat. We acknowledge support by the Austrian
Science Fund FWF through the project SFB F40.
We thank Klaus M\o lmer for his helpful remarks and open discussions.


\bibliography{aps_bib}

\end{document}